\documentstyle[12pt]{article}

\def\be{\begin{equation}}
\def\ee{\end{equation}}
\def\ben{\begin{displaymath}}
\def\een{\end{displaymath}}
\def\ba{\begin{array}{c}}
\def\ea{\end{array}}
\begin{document}

\titlepage
\vspace*{2cm}

\begin{center}{\Large \bf
The conditionally exactly solvable potentials:

A misunderstanding}\end{center}

\vspace{10mm}

\begin{center}
Miloslav Znojil
\vspace{3mm}

\'{U}stav jadern\'e fyziky AV \v{C}R, 250 68 \v{R}e\v{z},
Czech Republic\\

e-mail: znojil @ ujf.cas.cz

\end{center}

\vspace{5mm}

\section*{Abstract}

We detect an omission in the paper ``Conditionally exactly soluble
class of quantum potentials" by A. de Souza Dutra [Phys. Rev. A 47
(1993) R2435]. There, two strongly singular $s-$wave bound state
problems have been claimed completely solvable in closed form.
Unfortunately, all the displayed wave functions represented merely
asymptotically correct (so called ``Jost") solutions and did not
satisfy the appropriate threshold boundary condition. We show that
the incorporation of its standard form only leads to a very
partial exact solvability at a single energy and for special
couplings.

\newpage


\noindent For the two strongly singular $s-$wave potentials given,
in the units $\hbar = 2 \mu =1$, by the formulae
\be
V_1(r)={A \over r} + {B \over r^{1/2} } +{G \over r^2},
\ \ \ \ \ \ \ \ G = G_0 = -{3 \over 16},
\label{pota}
 \ee
and
\be
V_2(r)={A \, r^{2/3}} + {B \over r^{2/3} } +{G \over r^2},
\ \ \ \ \ \ \ \ G = g_0=
 -{5 \over 36}
\label{potbe}
  \ee
A. de Souza Dutra \cite{Dutra} offered the explicit elementary
wave functions as well as closed formulae for all their
bound-state energies. One of the three couplings is not free: This
entitled him to coin their ``conditionally" exactly soluble (CES)
status. In what follows we intend to demonstrate that in the sense
of the Ushveridze's monograph \cite{Ushveridze} both these forces
$V_{1,2}(r)$ only remain {\em partially} solvable at certain
specific values of the energies $E$ and couplings $B$.

Our present main point is that all the solutions presented in ref.
\cite{Dutra} still have to satisfy an appropriate and, for reasons
to be made understandable here, forgotten boundary condition in
the origin. Indeed, it is well known that for a central potential,
Schr\"{o}dinger equation $ -\triangle \Psi(\vec{r}) + V(|\vec{r}|)
\Psi(\vec{r})= E \Psi(\vec{r}) $ degenerates to an infinite set of
the ordinary (often called radial) decoupled differential
equations
 \be
 -\, \frac{{\rm d}^2}{{\rm d} r^2} \psi(r)
+\frac{\ell(\ell+1)}{r^2} \psi(r) + V(r) \psi(r)= E \psi(r), \ \ \
\ \ \ \ \ \ell = 0, 1, \ldots \label{SEr}
  \ee
for the separate angular-momentum components of the whole original
wave function. The Newton's excellent review \cite{Newton}
summarizes the details. {\em Under the assumption of the
analyticity of $V(r)$ in the origin} it shows that and why the
standard physical requirement of normalizability of bound states $
||\Psi(\vec{r})|| < \infty $ is {\em strictly} equivalent to the
integrability of their partial waves,
\be
\psi(r) \in L_2(0,\infty). \label{normaliz}
  \ee
For $\ell = 1, 2, \ldots$ the unphysical component
$\psi_{irregular}(r) \approx r^{-\ell}$ of the general threshold
solution of eq. (\ref{SEr}) is manifestly non-integrable near $r
\approx 0$.

In the $s-$wave with $\ell=0$ a more subtle argumentation is
needed \cite{Newton}. In practice, the subtlety is usually avoided
by the replacement of eq. (\ref{normaliz}) by the boundary
condition
\be
\lim_{r \to 0}\psi(r) = 0.
\label{kveak}
 \ee
Even when we solve the ordinary harmonic oscillator the latter
boundary condition in the origin offers a more straightforward
recipe for numerical calculations. Let us repeat: {\em for
analytic potentials}, eqs. (\ref{normaliz}) and (\ref{kveak}) are
equivalent but the proof \cite{Newton} of their equivalence
immediately fails for the ``very next" non-analytic $ V(r) \approx
G\,r^{-2} $, $r \approx 0$, say, in the Kratzer's solvable
phenomenological model \cite{Kratzer} with $G \neq 0$ etc. One
must re-analyze the whole quantization procedure anew, even for
harmonic oscillator at $G\to 0$ \cite{Fluegge}.

For all the similar singular forces with the finite limit in eq.
(\ref{SEr}),
 \ben
G =\lim_{r \to 0}\ r^2 V(r) \neq 0
 \een
we have to re-define the dominant singularity $\ell(\ell+1) + G =
{\cal L}({\cal L} +1) $. The new parameter $ {\cal L} =\sqrt{\left
(\ell+\frac{1} {2}\right )^2 + G } -\frac{1}{2} $ enters then the
modified threshold solutions $\psi_{regular}(r) \approx r^{{\cal
L}+1}$ and $\psi_{irregular}(r) \approx r^{-{\cal L}}$. The
irregular one is eliminated as manifestly violating the
normalizability (\ref{normaliz}) at ${\cal L} \geq 1/2$.

The latter bound means $G\geq 3/4$ in $s-$wave with $\ell=0$.
Below such a strength of repulsion the Hamiltonian {\em ceases to
be self-adjoint}. The conclusion is strongly counter-intuitive.
Mathematically, the problem is serious. First spotted and analyzed
by Case \cite{Case}, it means that at $G < 3/4$, the textbook
quantization of the Kratzer-like singular models {\em is not
unique at all}. A more detailed discussion may be found in the
literature (cf., e.g., \cite{Frank} or \cite{Reed}). In its light,
physics community currently accepts a unique way of quantization
which is, mathematically speaking, a mere regularization. It is
often supported by the various sufficiently robust {\it ad hoc}
arguments (cf., e.g., \cite{Fluegge} on pp. 157 and 167 or ref.
\cite{Stillinger}).

For our present purposes, in the physical language of textbook
\cite{Landau}, the correct recipe may be formulated as follows.

\begin{itemize}

\item
In the domain of a weak repulsion we distinguish between the
physical $\psi_{regular}(r) \approx r^{{\cal L}+1}$ and unphysical
$\psi_{irregular}(r) \approx r^{-{\cal L}}$. As long as both of
them remain normalizable, we impose an extra, {\em stronger}
boundary condition in the origin,
\be
\lim_{r \to 0}\psi(r) = 0, \ \ \ \ \ \ \ \ G \in (0, 3/4).
\label{weak}
 \ee
It coincides with (\ref{kveak}) but its mathematical meaning of a
convenient choice of the most plausible self-adjoint extension is
different.

\item
In the domain of weak attraction, both solutions
$\psi_{regular}(r) \approx r^{{\cal L}+1}$ and
$\psi_{irregular}(r) \approx r^{-{\cal L}}$ are compatible with
eq.~(\ref{weak}). In a sensible physical theory which
distinguishes between the two, the replacement of eq. (\ref{weak})
by an even stronger artificial constraint is needed,
\be
\lim_{r \to 0}\psi(r)/\sqrt{r} = 0, \ \ \ \ \ \ \ \ G \in
(-1/4,0).
\label{weakdva}
 \ee
\item
Below the lower bound $G \leq -1/4$ one cannot prevent the
spectrum from collapse by any means. Particles would definitely
fall in the origin.

\end{itemize}

\noindent We may summarize: In practice, bound state solutions of
the Schr\"{o}dinger differential eq. (\ref{SEr}) may be
constructed in two ways, namely,
\begin{itemize}

\item
{\bf  [RS]}
as the regular solutions $\psi_{regular}(r)$
constrained by the asymptotic normalizability condition
 \ben \psi_{regular}(R) = 0, \
\ \ \ \ \ R \to \infty; \label{Jost}
 \een

\item
{\bf  [JS]}
from the so called Jost solutions $\psi_{Jost}(r)$,
{\em always} exhibiting the square-integrable asymptotic decrease
by definition.

\end{itemize}

\noindent The former regular-solution approach [RS] proves useful
within the framework of the standard Taylor series method
\cite{Ince} and in non-numerical context \cite{chov}.
Schr\"{o}dinger equation (\ref{SEr}) becomes converted into the
exactly solvable two-term recurrences at $q=0$ (harmonic
oscillator), into the three-term recurrences at $q=1$ (sextic
forces) etc \cite{Classif}. Rather unexpectedly, for all the
positive integers $q=1, 2, \ldots$, a few bound states may still
appear in an exact polynomial (i.e., terminating Taylor-series)
form. An explicit construction of these exceptional elementary
states is based on the solution of the Magyari's nonlinear
algebraic equations \cite{Magyari}. They determine a few energy
levels exactly and restrict also the free variability of the
available couplings.

In the latter context, potentials $V_{1,2}(r)$ exhibit a certain
incomplete dynamical symmetry and play an exceptional role as
quasi-exactly solvable in a certain narrower sense (cf. ref.
\cite{Ushveridze} for more details). This would make the ambitious
conclusions of ref. \cite{Dutra}, if they were all true, even more
important.

Their analysis must be based on the alternative option [JS] which
requires the threshold boundary condition (\ref{kveak}),
(\ref{weak}) or (\ref{weakdva}) \cite{Frank}. This is a core of
our present message. For the particular forces (\ref{pota}) and
(\ref{potbe}) such an approach has already thoroughly been tested
numerically in ref. \cite{Stillinger}. The Liouvillean
\cite{Liouville} change of variables $r\to x = r^{const}$ and
$\psi(r) \to x^{const} \chi(x)$ has been employed there. As long
as it leaves the form of the Schr\"{o}dinger equation unchanged,
it reduces all the bound-state problems with forces of the type
(\ref{pota}) and (\ref{potbe}) to their ``canonical" equivalents
with polynomial potentials
 \be
V_T(x) =a r^{-2}+{b \, r^2 } +{c\, r^4} + \ldots + y\,r^{4q} +
z\,r^{4q+2}, \ \ \ \ \ \ \ \ \ \ \ a > -1/4. \label{potaz}
 \ee
On this basis, we may easily deduce the leading-order solutions
(near the origin) also for our singular potentials $V_{1,2}(r)$ of
eqs. (\ref{pota}) and (\ref{potbe}),
 \ben
 \psi_{1,regular}(r) \sim r^{3/4},\ \ \ \ \ \ \ \ \ \
 \ \ \ \ \ \psi_{1,irregular}(r) \sim
r^{1/4}, \label{nn9}
 \een
 \ben
\psi_{2,regular}(r) \sim r^{5/6}, \ \ \ \ \ \ \ \ \ \ \ \ \ \ \ \
\psi_{2,irregular}(r) \sim r^{1/6}.
 \een
This is to be compared with the Dutra's wave functions: Say, for
potential $V_1(r)$ we may quote equation Nr. (9) from
\cite{Dutra},
 \be \psi_1^{(D)}(r) = C\,r^{1/4}
\exp\left [- \frac{1}{2} \beta^2\, \left (
 r^{1/2}-\frac{B}{2E}\right )^2\right ]
 \ H_n \left [\beta\, \left (
r^{1/2}-\frac{B}{2E}\right )\right ]. \label{n9}
 \ee
Its energies $E = -\beta^4/4$ are parametrized by $\beta=\beta_n$
and numbered by an integer $n=0, 1, \ldots$. The formula also
contains a certain normalization constant $C = C_n$ and Hermite
polynomials $H_n(x)$. We immediately detect an inconsistency of
the latter two equations near the origin.

Similar observation is also made for $\psi_2(r)$ from equation Nr.
(13) in \cite{Dutra}: None of the Dutra's wave functions satisfies
the physical boundary condition (\ref{weakdva}). An explanation of
this obvious misunderstanding is in fact not too difficult: The
solutions were merely constrained by the too weak (though, in
practice, much more frequently encountered) and, hence,
inapplicable threshold condition (\ref{kveak}). We may summarize
that the inconsequent use of the boundary conditions would lead to
a physically absurd spectrum covering the whole real line, $E \in
(-\infty,\infty)$.

The Dutra's ``non-anonymous" (i.e., Hermite-polynomial) solutions
have already evoked a non-negligible response in the current
literature. As an example one might quote the paper \cite{Nag}.
Its authors relied on the physical correctness of the Dutra's
argumentation and were misguided in their mathematical
appreciation of the role of supersymmetry in the similar problems.
Still, the majority of their argument remains valid. Hence, let us
show, in the conclusion, how the Dutra's exceptional solutions
could be ``saved" for similar applications.

Obviously, one has to incorporate simply the necessary constraint
(\ref{weakdva}). An inspection, say, of our sample eq. (\ref{n9})
reveals that $\psi^{(D)}_1(r)$ satisfies condition (\ref{weakdva})
{\em if and only if} its Hermite-polynomial component acquires an
exact nodal zero in the origin. In terms of the known numbers
$X=X(n,k)$ (calculated as the $k-$th nontrivial zeros of $H_n(X)$,
cf. Table~1) this requirement, unfortunately, fixes the
non-Coulombic coupling as a function of the energy $E =
-\beta^4/4$,
 \be
B = \frac{1}{2}\,X\, \beta^3 \neq 0. \label{Be}
 \ee
This makes both these values coupled to the additional (in fact,
Magyari's \cite{Magyari}) constraint. As a cubic equation for the
energy $E$ it appears under Nr. (8) in ref. \cite{Dutra}. This
algebraic selfconsistency condition must be combined with eq.
(\ref{Be}). The resulting polynomial equation in $\beta$ (of
twelfth degree!) is easily factorized in closed form. Its real
roots we need are
 \ben
 \beta=\beta(n, k) = 2\, \sqrt{\frac{-A}
{2n +1-X^2(n,k) }
 }.
 \een
They all exist for any $A < 0$. This is an important conclusion:
let us note that eq. Nr. (8) of ref. \cite{Dutra} re-appears as
equation Nr. (16) in ref. \cite{Nag}, etc.

For illustration, let us finally fix the scale $A=-1$ and display
the first few non-numerical specifications of energies
$E=-\beta^4$ and their couplings (\ref{Be}) in Table~2. The same
parameters are to be used also in the definition (\ref{n9}) of the
correct bound-state wave function. {\it Mutatis mutandis}, the
entirely parallel ``return to validity" applies also to
$\psi_2^{(D)}(r)$ in \cite{Dutra}. We omit the details here,
re-emphasizing only that both the Dutra's expressions
$\psi^{(D)}_{1,2}(r)$ are elementary and still satisfy the
Schr\"{o}dinger differential equation, exhibiting also the correct
asymptotic behaviour. Thus, we may return, say, to the paper by
Dutt et al \cite{Khare}, originally motivated by ref. \cite{Dutra}
as well. In the light of our present notes, the importance of the
latter paper increases: Its authors have, involuntarily, found and
constructed {\em the first} CES example in one dimension!


\section*{Acknowledgements}

Years long discussions of the subject with my colleagues in Theory
Group of NPI in \v{R}e\v{z} and with authors of refs. \cite{Nag}
and \cite{Khare} contributed to this paper. An anonymous referee
attracted my attention {\it ad fontes}  \cite{Case} and
\cite{Frank}. The reference to the highly relevant paper
\cite{Stillinger} was kindly communicated to me by A. de Souza
Dutra in his non-anonymous referee report. He also informed me
about his correspondence with F. H. Stillinger, the subsequent
private communication with whom is also acknowledged.

\newpage
\begin{center}
Table 1. Non-vanishing zeros $X=X(n,k)$

of the first few  Hermite polynomials $H_n(X)$.
\end{center}

$$ \begin{array}{||cc|cccc||} \hline  \hline
 &k & 4
 &3&2 &1 \\
 n& & & & & \\
 \hline
0 & &- &- &- &- \\
1 & &- &- &- &- \\
2&& -&-\sqrt{ 1/2}
& \sqrt{ 1/2}&-
 \\
3&& -&-\sqrt{ 3/2}
& \sqrt{ 3/2}&-
 \\
  4& & -\sqrt{(3 + \sqrt{6})/2}&
   -\sqrt{(3 - \sqrt{6})/2}&
   \sqrt{(3 - \sqrt{6})/2}&
   \sqrt{(3 + \sqrt{6})/2}\\
  5 && -\sqrt{(5 + \sqrt{10})/2}&
   -\sqrt{(5 - \sqrt{10})/2}&
   \sqrt{(5 - \sqrt{10})/2}&
   \sqrt{(5 + \sqrt{10})/2}\\
\hline \hline
\end{array}
$$

\vspace{1cm}
\begin{center}
 Table 2. Parameters of the first few simplest
quasi-exact states in $V_1(r)$.

$M$ counts nodes in $\psi_1(r)$: $M=0$ means
ground state, etc. \end{center}

$$
\begin{array}{||ccc||cr|cr||} \hline  \hline
 M & n&k& \multicolumn{2}{c|}{ {\rm fixed\
coupling}\  B'=B/8} & \multicolumn{2}{c||}{ {\rm binding\ energy}
\ E}\\ \hline \hline
 0&2&1&  \sqrt{1/9^3}&\sim 0.0370
& -(4/9)^2 &\sim -0.197 \\
 &3&1&  \sqrt{3/11^3}&\sim 0.0475
& -(4/11)^2 &\sim -0.055 \\
 & 4& 1&
  \sqrt{(3+ \sqrt{6})/(15- \sqrt{6})^3}
&\sim 0.0525& -[4/(15+ \sqrt{6})]^2 &\sim -0.029 \\
 & 5& 1&
 \sqrt{(5+ \sqrt{10})/(17- \sqrt{10})^3}
  &\sim 0.0555& -[4/(17+ \sqrt{10})]^2 &\sim -0.018 \\
\hline
 1& 2& 2&  -\sqrt{1/9^3}
&\sim- 0.037& -(4/9)^2 &\sim -0.132 \\
 & 4& 2&
  \sqrt{(3- \sqrt{6})/(15+ \sqrt{6})^3}
&\sim 0.0102& -[4/(15- \sqrt{6})]^2 &\sim -0.047  \\
 & 5& 2&
  \sqrt{(5- \sqrt{10})/(17+ \sqrt{10})^3}
& \sim 0.0150& -[4/(17- \sqrt{10})]^2 &\sim -0.028 \\ \hline
 2& 3& 2&  -\sqrt{3/11^3}
&\sim- 0.047&-(4/11)^2 &\sim -0.055 \\
 & 4& 3&-\sqrt{(3- \sqrt{6})/(15+ \sqrt{6})^3}
& \sim -0.010& -[4/(15- \sqrt{6})]^2 &\sim -0.047 \\ \hline
 3& 4& 4& -\sqrt{(3+ \sqrt{6})/(15- \sqrt{6})^3}
& \sim -0.053& -[4/(15+ \sqrt{6})]^2 &\sim -0.029 \\
 & 5& 3& -\sqrt{(5- \sqrt{10})/(17+ \sqrt{10})^3}
&\sim -0.015& -[4/(17- \sqrt{10})]^2 &\sim -0.028  \\ \hline
 4& 5& 4&
-\sqrt{(5+ \sqrt{10})/(17- \sqrt{10})^3}&\sim -0.056& -[4/(17+
\sqrt{10})]^2 &\sim -0.039\\
 \hline \hline
\end{array}
$$

\end{document}